\documentstyle[aps,prb,twocolumn,epsf,graphics,psfig]{revtex}

\begin{document}
\twocolumn[\hsize\textwidth\columnwidth\hsize\csname @twocolumnfalse\endcsname
\title{Curie Temperature Trends in (III,Mn)V Ferromagnetic Semiconductors}
\author{T. Jungwirth$^{1,2}$, J\"{u}rgen K\"{o}nig$^{3}$, Jairo Sinova$^{1}$,
J. Ku\v{c}era$^2$, and A.H. MacDonald$^{1}$}
\address{$^{1}$Department of Physics,
The University of Texas at Austin, Austin, TX 78712 \\}
\address{$^{2}$Institute of Physics ASCR, Cukrovarnick\'a 10,
162 53 Praha 6, Czech Republic}
\address{$^{3}$Institut f\"{u}r Theoretische Festk\"{o}rperphysik,
Universit\"{a}t Karlsruhe, D-76128 Karlsruhe, Germany
}
\date{\today}
\maketitle
\begin{abstract}
We present a theoretical survey of ferromagnetic transition
temperatures in cubic (III,Mn)V semiconductors based on a model
with $S=5/2$ local moments exchange-coupled to itinerant holes 
in the host semiconductor valence band.  Starting from the simplest
mean-field theory of this model, we estimate the $T_c$ enhancement 
due to exchange and correlation in the itinerant-hole system,
and the $T_c$ suppression due to collective fluctuations of the ordered moments.
We show that high critical temperatures in these ferromagnetic
semiconductors require both the large magnetic susceptibility contribution
from the valence band's heavy holes and the large spin stiffness that results 
from strong spin-orbit coupling in the valence band.  Our calculations demonstrate that 
this model for the ferromagnetism of these systems 
is fully consistent with the room-temperature ferromagnetism
reported for Mn doped nitrides.
\end{abstract}
\pacs{75.50.Pp,75.30.Gw,73.61.Ey}

\vskip2pc]

The 1992 discovery\cite{ohnoprl92}
of hole-mediated ferromagnetic order in (In,Mn)As has
motivated research\cite{ohnoapl96} on Mn doped GaAs and other III-V host materials.
Ferromagnetic transition temperatures\cite{ohnosci98} in excess of 100~K
and long spin-coherence times\cite{kikkawanature99} in GaAs
have fueled hopes that a new magnetic medium is emerging that could open new
pathways for information processing and storage technologies. 
Recently, confirmation\cite{reedapl01}
of the room-temperature ferromagnetism predicted\cite{dietlsci00,dietlprb01} 
in (Ga,Mn)N has added to interest in this class of materials.

When substituted on the cation site of a III-V semiconductor, 
it is generally accepted that Mn acts as an acceptor
leaving a Mn$^{2+}$  half-filled d-shell with angular momentum
$L=0$ and spin $S=5/2$. There is also  experimental 
evidence\cite{szczytkoprb99,linnarssonprb97,okabayashi,ohnojmmm99}
that ferromagnetism occurs in these materials because of interactions between Mn local
moments that are mediated by holes in the semiconductor valence band.
The local moments and the itinerant holes are coupled 
in this approach by an antiferromagnetic exchange interaction. 
The participation of itinerant holes in the ferromagnetism
of these diluted magnetic semiconductors 
adds to their richness, leading to strong magnetotransport effects 
that might have important applications and, of particular interest to us here,
to transition temperatures that are 
sensitive to the details of the host band structure. 

The description of  ordered states in (III,Mn)V systems is greatly simplified
by a virtual-crystal type approximation in which
the Mn ion distribution is replaced by a continuum with the same 
impurity density.\cite{dietlsci00,dietlprb97,jungwirthprb99,bookchapter}  
Disorder effects associated with randomness in the Mn ion sites can
then be treated perturbatively, when necessary.  (For example randomness in 
the Mn distribution likely limits the conductivity at low-temperatures.) 
This strategy will fail in the limit of dilute Mn ions and 
also when the exchange interaction between band and Mn spins 
is too strong.\cite{schliemannprb01}
It does, however, seem to be reliable in the limit of principal 
interest, that of high Mn densities and high critical temperatures,
where the holes are metallic and their interaction with Mn acceptors will be 
effectively screened.  

There is at present considerable activity directed toward the growth of 
many different (III,V) materials containing Mn. 
In this paper we present and discuss theoretical predictions\cite{msweb}
for their ferromagnetic critical temperatures based on the continuum
model.  We go beyond the standard mean-field theory of this model by 
accounting for the role of Coulomb interactions holes in the valence
band, which enhances the critical temperature, and for correlations in 
Mn ion orientations which reduce the energetic cost of unaligned spin
configurations and therefore lower the critical 
temperature.\cite{konigprl00,schliemannapl01,konigprb01}

When disorder and hole correlations are 
neglected, the mean-field transition temperature of this model is 
given by\cite{jungwirthprb99,dietlsci00}
\begin{equation}
  k_B T_c = \frac{N_{\rm Mn} S (S+1)}{3} 
  \frac{J_{\rm pd}^2\chi_f}{(g \mu_B)^2} \; ,
\label{tc}
\end{equation}
where $N_{\rm Mn}=4x/a_{\rm lc}^3$ is the Mn density in Mn$_x$III$_{1-x}$V
zinc-blende semiconductors with a lattice constant $a_{\rm lc}$, $J_{\rm pd}$
is the localized spin -- itinerant spin exchange 
coupling constant, and $\chi_f$ is
the itinerant hole magnetic susceptibility.
To understand the qualitative implications of this $T_c$-equation~(\ref{tc}),
we first consider the case of a model itinerant electron 
system with a single spin-split band and an effective mass $m^{\ast}$.
The kinetic energy of the band holes gives a contribution to
the susceptibility:
\begin{equation}
  \frac{\chi_f^{\rm kin}}{(g\mu_B)^2}=\frac{m^{\ast}k_F}{4\pi^2\hbar^2}
  \; ,
\label{etk}
\end{equation}
where $k_F$ is the Fermi wavevector. For weak interactions, the exchange energy of the
spin-polarized parabolic-band model adds a contribution: 
\begin{equation}
  \frac{\chi_f^{\rm ex}}{(g\mu_B)^2}=
  \frac{e^2 (m^{\ast})^2}{4\pi^3\varepsilon \hbar^4} \; ,
\label{etx}
\end{equation}
where $\varepsilon$ is the dielectric constant of the host semiconductor.
At high hole densities $p$, the kinetic-energy term dominates and 
the mean-field critical temperature, $T_c^{\rm MF}$,
is proportional to the Fermi wavevector, i.e., to $p^{1/3}$. 
Equations~(\ref{etk}) and (\ref{etx})
also show that the band contribution to the
susceptibility increases linearly with $m^{\ast}$
while the exchange correction is proportional to
$(m^{\ast})^2$. 
(Using this simple band model as a guide, hole correlation effects, which will not be
discussed in detail here, suppress the critical temperature by several percent for typical 
experimental hole densities, $p\sim 0.1$~nm$^{-3}$.)

\begin{figure}
\epsfxsize=3.3in
\hspace*{-0.5cm}
\centerline{
\epsffile{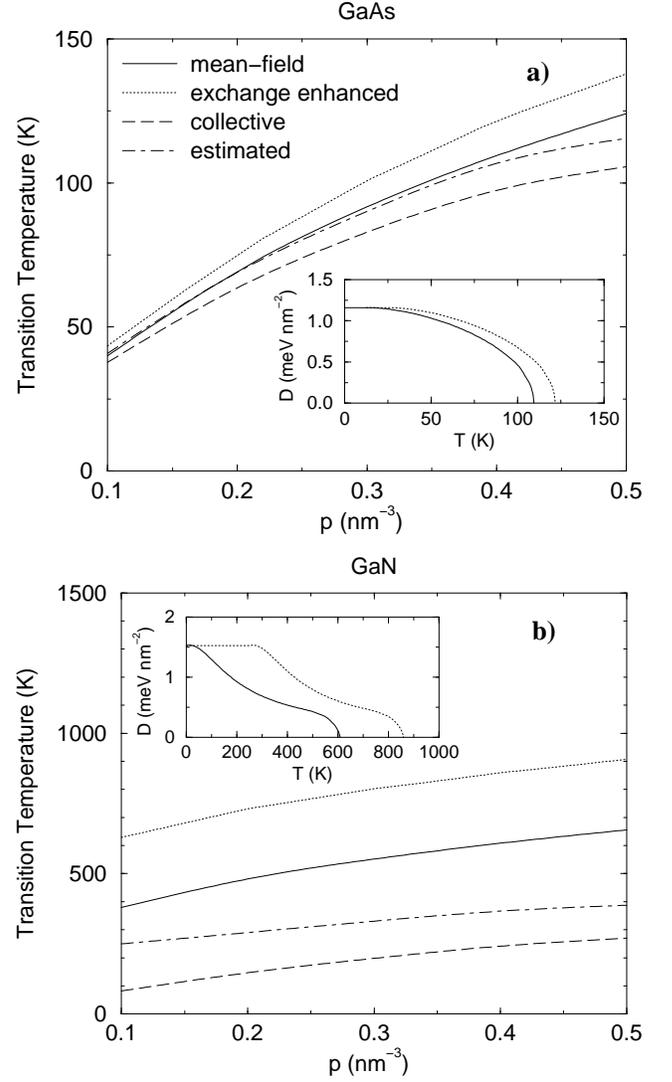}}

\vspace*{.0cm}

\caption{Main graphs:
Mean-field ($T_c^{\rm MF}$), exchange enhanced ($T_c^{\rm ex}$),
collective ($T_c^{\rm coll}$), and estimated ($T_c^{\rm est}$) ferromagnetic
transition temperatures in GaAs (a) and GaN (b) host semiconductors
doped with 5\% of Mn are plotted as a function of itinerant-hole density.
Insets: temperature dependence of the spin stiffness (solid lines)
calculated from
the zero-temperature stiffness and from the mean-field magnetizations.
The hole density $p=0.4$~nm$^{-3}$ and $x=5$\% for both
GaAs (a) and GaN (b) hosts. Shifted (dotted) curves account for local
magnetization enhancement due to exchange interaction in the itinerant hole
system.
}
\label{gaasn}
\end{figure}

To obtain quantitative predictions for the critical temperature, 
it is necessary to evaluate the kinetic and exchange contribution to the itinerant hole
susceptibility using a realistic six-band Kohn-Luttinger 
model\cite{dietlsci00,dietlprb01,abolfathprb01},
instead of the parabolic-band model.  The band Hamiltonian
contains the spin-orbit splitting parameter $\Delta_{\rm so}$ and
three other phenomenological parameters, $\gamma_1$, $\gamma_2$, and $\gamma_3$,
whose values for the specific III-V host can be found, e.g., 
in Refs.~\onlinecite{dietlprb01,iiiv}.  Only zinc-blende crystals are
considered here; in nitrides, which crystallize in both
zinc-blende and wurtzite structures, the critical temperature is
expected to be only weakly dependent on the lattice configuration.\cite{dietlprb01}
The non-interacting-hole mean-field, $T_c^{\rm MF}$, and the 
exchange-enhanced, $T_c^{\rm ex}$, critical temperatures
for GaAs and GaN host semiconductors doped with 5\% of Mn are plotted 
in Fig.~\ref{gaasn} as a function of the hole density. Transition
temperatures for other III-V hosts and for hole densities $p=0.1$ and 0.5~nm$^{-3}$ are listed in 
Table~\ref{tciiivs}. We have assumed $J_{\rm pd}=55$~meV~nm$^{-3}$ for all
III-V hosts.\cite{dietlprb01,ohnojmmm99}  
In the density range considered, only the two heavy-hole and two light-hole 
bands are occupied in arsenides and antimonides.  However, the mixing
between these four bands and the two spin-orbit split-off bands is
strong and must be accounted for. In nitrides and phosphides, spin-orbit
coupling is weaker and all six bands are occupied by holes.
The numerical results for the transition temperature are consistent with the qualitative 
analysis based on the parabolic band model:  
$T_c^{\rm MF}$ follows roughly the $p^{1/3}$ dependence, the $\sim 10-50$\% exchange
enhancement of the critical temperature has a weaker density dependence.

For the thoroughly studied (Ga,Mn)As material\cite{ohnojmmm99} doped with 
5\% of Mn and with $p=0.35$~nm$^{-3}$, we find $T_c^{\rm MF}=102$~K and
$T_c^{\rm ex}=112$~K. This result is in remarkably good agreement
with the experimental transition temperature of 110~K . Good
agreement between experiment\cite{ohnojmmm99} and
mean-field theory is also found in (In,Mn)As compounds,
that have lower critical temperatures compared to their (Ga,Mn)As
counterparts, primarily because they have smaller band masses.
An especially important system is (Ga,Mn)N, which appears to exhibit 
room-temperature 
ferromagnetism.\cite{reedapl01} Our calculations
for this III-V semiconductor give $T_c^{\rm MF}\approx
400$-600~K and $T_c^{\rm ex}\approx 600$-900~K for Mn content 
$x=5$\% over the range of hole densities studied.
Although we cannot compare those numbers directly to 
the recently measured $T_c=228-370$~K in (Ga,Mn)N
samples\cite{reedapl01}, since they have  
inhomogeneous doping profiles (resulting from the chemical vapor
deposition growth technique),  
the mean-field theory predictions have the correct order of magnitude but
appear to overestimate the transition temperature.

We now establish the quantitative reliability of the mean-field theory
in arsenides (and also in antimonides and lower-carrier-density
phosphides) and explain
why mean-field theory overestimates the transition temperature in
the heavy effective mass nitrides. Our arguments are based on the estimates
of the suppression of the transition temperature due to 
collective spin-wave fluctuations.
We emphasize below that mean-field theory would fail badly in any of these materials
if the appropriate band model were really a single parabolic band with a 
mass equal to the material's heavy-hole mass.  Strong spin-orbit coupling
in the valence band, which yields band states whose orbital and spin
character changes completely as the band wave-vector varies over the 
Brillouin zone,\cite{gergely0108477}  enhances the spin stiffness, increases the energy 
of correlated reduced magnetization configurations, and reduces
the importance of corrections to the mean-field-theory result.\cite{konigprb01}  

Isotropic ferromagnets have spin-wave Goldstone collective modes whose 
energies vanish at long wavelengths,
\begin{equation}
\label{long-wavelength}
   \Omega_{k} = D k^2 + {\cal O} (k^4)\, ,
\end{equation}
where $k$ is the wavevector of the mode.
Spin-orbit coupling breaks rotational symmetry and leads to a finite gap.
According to our numerical studies,\cite{konigprb01} this gap is small however, much smaller
than $T_c$ for example, and plays a role in magnetic fluctuations only at 
very low temperatures.  Spin-wave excitations reduce the total spin by one,
at an energy cost that is, at least at long wavelengths,
much smaller than the mean-field value, $J_{\rm pd} s^{(0)}$ where $s^{(0)}$ is the 
mean-field band spin-density.  The importance of these correlated spin excitations,
neglected by mean-field theory, can be judged by evaluating an approximate 
$T_c$ bound based on the following argument which uses a Debye-like model
for the magnetic excitation spectrum.
When spin-wave interactions are neglected, 
the magnetization vanishes at the temperature where the number of excited spin 
waves equals the total spin of the ground state:  
\begin{equation}
   N_{\rm Mn}S = {1\over 2 \pi^2} \int_0^{k_D} dk\, k^2 n(\Omega_{k}) \, ,
\end{equation}
where $n(\Omega_{k})$ is the Bose occupation number and the Debye cutoff,
$k_D=(6\pi^2N_{\rm Mn})^{1/3}$. 
It follows that the ferromagnetic transition temperature cannot exceed 
\begin{equation}
   k_BT_c^{\rm coll} = {2S+1\over 6}  k_D^2 D(T_c^{\rm coll}) \; .
\label{collective_tc}
\end{equation}
In applying this formula to estimate $T_c$ we have 
approximated the temperature dependence of the spin stiffness by 
\begin{equation}
D(T)=D_0\langle S\rangle(T)/S\; ,
\label{dt}
\end{equation}
as we did in Ref.~\onlinecite{dietlrapid01},
where $D_0$ is the zero-temperature stiffness,\cite{konigprb01} 
and $\langle S\rangle(T)$
is the mean-field Mn polarization\cite{abolfathprb01} at a temperature $T$.
If the difference between $T_c^{\rm coll}$ and $T_c^{\rm MF}$ is large, 
the typical local valence-band carrier polarization will remain
finite above the critical temperature and 
ferromagnetism will disappear only because of the loss of long-range spatial order,
the usual circumstance for transition metal ferromagnetism for example.

In discussing corrections to mean-field-theory $T_c$ estimates, 
we compare spin-stiffness results obtained with the simple two-band
and realistic six-band models. Details on the formalism used to calculate
$D_0$ can be found in Refs.~\onlinecite{konigprl00,konigprb01}.
We find that the zero-temperature 
spin stiffness is always much larger in the six-band model. 
For (Ga,Mn)As, for example, the two-band model underestimates
$D_0$ by a factor of $\sim$10-30 over the range of
hole densities considered.  Furthermore, the trend is 
different: in the two-band model the stiffness decreases with 
increasing density, while for the six-band description the initial
is followed by a saturation.  Even in the limit of low carrier concentrations,
it is not only the (heavy-hole) mass of the lowest band which is important for 
the spin stiffness.  In the realistic band model, heavy-holes have
their spin and orbital angular momenta aligned approximately along the 
direction of the Bloch wavevector.  Exchange interactions with Mn spins
mix the heavy holes with more dispersive light holes.
Our calculations show that heavy-light mixing is responsible for the 
relatively large spin stiffnesses which are responsible for the 
general success of mean-field theory.  {\em Crudely, the large mass 
heavy-hole band dominates the spin susceptibility and enables local magnetic
order at high temperatures, while the dispersive light-hole band
dominates the spin stiffness and enables long-range magnetic order.} The 
multi-band character of the semiconductor valence band plays an essential role 
in the ferromagnetism of these materials.

In the insets of Figs.~\ref{gaasn}(a) and (b), 
solid lines show the spin stiffness for GaAs and GaN, doped with 5\%
of Mn and with $p=0.4$~nm$^{-3}$, 
as a function of temperature.  The critical temperatures, $T_c^{\rm coll}$,
were obtained from these curves and from 
Eqs.~(\ref{collective_tc}) and (\ref{dt}). $T_c^{\rm coll}$, 
calculated using a non-interacting hole approximation, and
$T_c^{\rm ex}$, calculated including exchange interactions but based on a mean-field
treatment of the Mn spin-orientation thermodynamics, 
represent approximate lower and upper bound for the ferromagnetic transition
temperature in (III,Mn)V semiconductors. 
{\em With the exception of nitrides and higher-carrier-density phosphides, 
exchange enhancement and
correlated fluctuations change $T_c$ by less than 
20\%, and these two influences will tend to cancel.  This property
explains the remarkable success of the mean-field theory
in these itinerant ferromagnets.
In the heavy effective mass nitrides, mean-field theory looses
its quantitative accuracy, but still predicts the correct order
of magnitude for the ferromagnetic transition temperature.}

The dot-dashed lines in Fig.~\ref{gaasn}, and Table~\ref{tciiivs} summarize 
critical temperature estimates $T_c^{\rm est}$ obtained
from Eqs.~(\ref{collective_tc}) and (\ref{dt}) and
by shifting the $D(T)$-curves in the insets of Fig.~\ref{gaasn} along
$x$-axis by $T_c^{\rm ex} - T_c^{\rm MF}$. This approach accounts, in 
an approximate way, for both exchange interactions 
and correlated spin fluctuations.  
For arsenides, antimonides, and lower-carrier-density
phosphides, $T_c^{\rm est}$ is very
close to $T_c^{\rm MF}$ justifying the mean-field description of
ferromagnetism in these semiconductor compounds. The estimated
critical temperatures for (Ga,Mn)N, $T_c^{\rm est}=200-400$~K,  reflect a more
significant suppression of ferromagnetism due to collective excitations.
Our theoretical $T_c^{\rm est}$ are in good agreement with  available 
experimental transition temperatures  in studied (III,Mn)V diluted magnetic
semiconductors, including (Ga,Mn)N.

The work was supported by the DARPA/ONR
Award No. N00014-00-1-095, the Welch Foundation, NSF under grant DMR-0115947, 
the Deutsche Forschungsgemeinschaft under the Emmy-Noether program 
KO 1987/2-1, the EU COST program, the Grant Agency of the Czech Republic under 
grant 202/02/0912, and by the Ministry of Education of the Czech Republic 
under grant OC P5.10.


\begin{table}
\begin{center}   
\begin{tabular}{cccccc}
host & $p$ (nm$^{-3}$)
& $T_c^{\rm MF}$  & $T_c^{\rm ex}$ & $T_c^{\rm coll}$ & $T_c^{\rm est}$ (K)
\\ \hline
AlAs & 0.1 & 45 & 53 & 41 & 47\\
     & 0.5 & 134 & 158 & 105 & 119\\ \hline
GaAs & 0.1 & 40 & 43 & 38 & 41\\
     & 0.5 & 124 & 138 & 106 & 115\\ \hline
InAs & 0.1 & 14 & 15 & 14 & 15\\
     & 0.5 & 41 & 44 & 40 & 41\\ \hline
AlSb & 0.1 & 19 & 22 & 18 & 20\\
     & 0.5 & 58 & 64 & 49 & 53\\ \hline
GaSb & 0.1 & 18 & 19 & 18 & 19\\
     & 0.5 & 85 & 88 & 82 & 85 \\ \hline
InSb & 0.1 & 11 & 12 & 11 & 11 \\
     & 0.5 & 37 & 38 & 35 & 36 \\ \hline 
AlP  & 0.1 & 94 & 127 & 73 & 94\\ 
     & 0.5 & 173 & 218 & 105 &121\\ \hline
GaP  & 0.1 & 57 & 70 & 50 & 60 \\
     & 0.5 & 101 & 115 & 43 & 45 \\ \hline
InP  & 0.1 & 66 & 80 & 60 & 70 \\
     & 0.5 & 136 & 163 & 103 & 118 \\ \hline
GaN  & 0.1 & 379 & 629 & 81 & 250 \\
     & 0.5 & 656 & 907 & 270 & 387 \\ \hline  
InN  & 0.1 & 308 & 549 & 89 & 240 \\
     & 0.5 & 531 & 777 & 303 & 423 \\  
\end{tabular}
\end{center}
\caption{
Mean-field ($T_c^{\rm MF}$), exchange enhanced ($T_c^{\rm ex}$),
collective ($T_c^{\rm coll}$), and estimated ($T_c^{\rm est}$) ferromagnetic
transition temperatures in III-V host semiconductors
doped with 5\% of Mn and with itinerant hole densities
$p=0.1$ and $p=0.5$~nm$^{-3}$.
}
\label{tciiivs}
\end{table}

\vspace{-.5cm}

\begin{figure}
\epsfxsize=3.3in
\hspace*{-0.5cm}
\centerline{
\epsffile{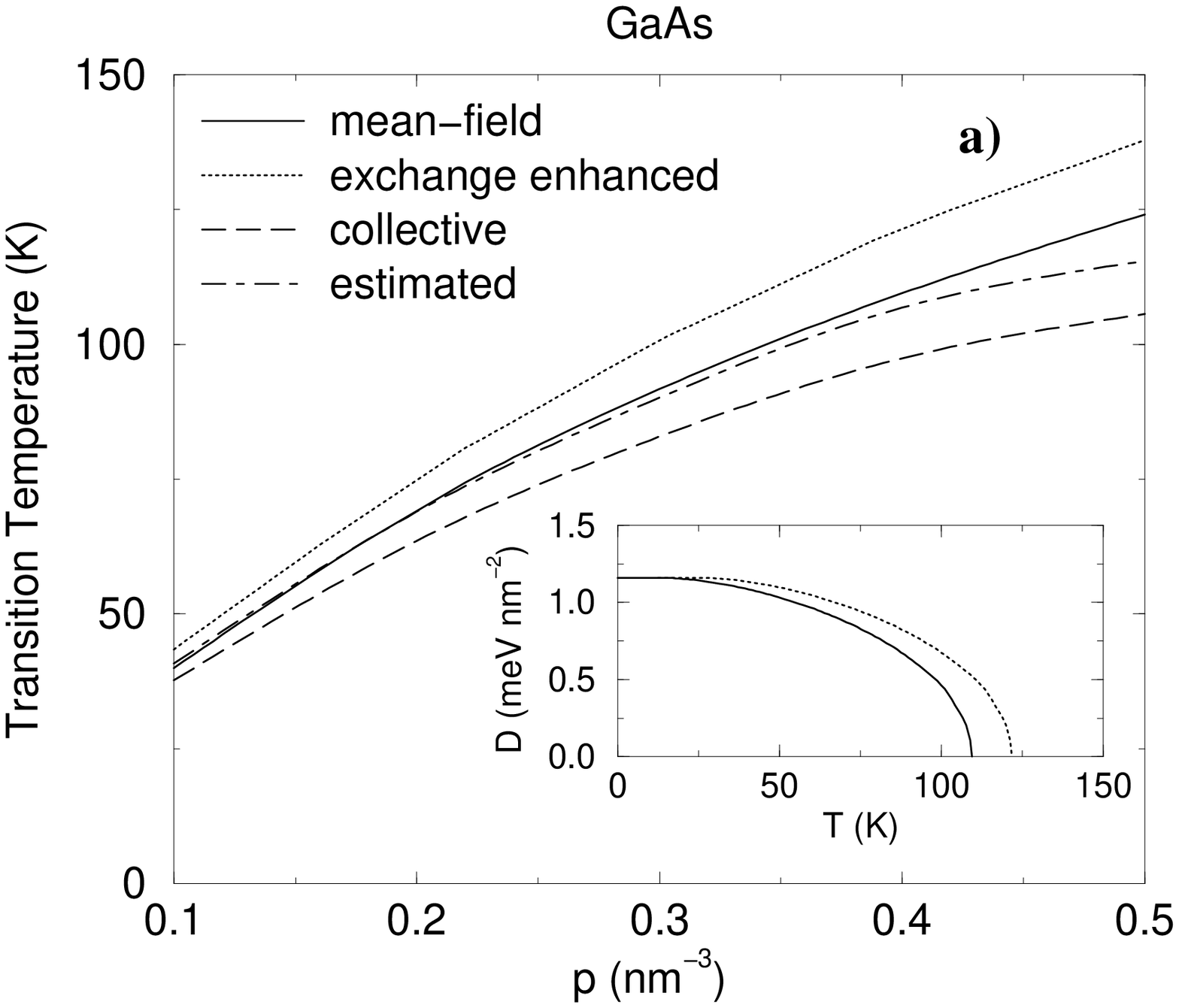}}

\vspace*{.7cm}

\end{figure}

\vspace{-1.5cm}

\begin{references}
\bibitem{ohnoprl92}
H. Ohno, H. Munekata, T. Penney, S. von Moln\'{a}r, and L.L. Chang,
Phys. Rev. Lett. {\bf 68}, 2664 (1992).

\bibitem{ohnoapl96}
H. Ohno, A. Shen, F. Matsukura, A. Oiwa, A. Endo, S. Katsumoto, and Y. Iye,
Appl. Phys. Lett. {\bf 69}, 363 (1996).

\bibitem{ohnosci98}
H. Ohno, Science {\bf 281}, 951 (1998);
F. Matsukura, H. Ohno, A. Shen, and Y. Sugawara,
Phys. Rev. B {\bf 57}, R2037 (1998).

\bibitem{kikkawanature99}
J.M. Kikkawa and D.D. Awschalom, Nature {\bf 397}, 139 (1999);
D.D. Awschalom and N. Samarth,
J. Magn. Magn. Mater. {\bf 200}, 130, (1999);
I. Malajovich, J.J. Berry, N. Samarth, and D.D. Awschalom,
Nature {\bf 411}, 770 (2001).


\bibitem{reedapl01}
M.L. Reed, N.A. El-Masry, H.H. Stadelmaier, M.K. Ritums,
M.J. Reed, C.A. Parker, J.C. Roberts, and S.M. Bedair,
Appl. Phys. Lett. {\bf 79}, 3473 (2001).

\bibitem{dietlsci00}
T. Dietl, H. Ohno, F. Matsukura, J. Cibert, and D. Ferrand, Science
{\bf 287}, 1019 (2000).

\bibitem{dietlprb01}
T. Dietl, H. Ohno, and F. Matsukura,
Phys. Rev. B {\bf 63}, 195205 (2001).

\bibitem{szczytkoprb99}
J. Szczytko, A. Twardowski, K. \'Swiatek, M. Palczewska, M. Tanaka,
T. Hayashi, and K. Ando,
Phys. Rev. B {\bf 60}, 8304 (1999).

\bibitem{linnarssonprb97}
M. Linnarsson, E. Janz{\'e}n, B. Monemar, M. Kleverman, and A. Thilderkvist,
Phys. Rev. B {\bf 55}, 6938 (1997).

\bibitem{okabayashi}
J. Okabayashi, A. Kimura, O. Rader, T. Mizokawa, A. Fujimori, T. Hayashi,
and M. Tanaka, Phys. Rev. B {\bf 58}, R4211 (1998);
{\em ibid} {\bf 64}, 125304 (2001).

\bibitem{ohnojmmm99}
H. Ohno, J. Magn. Magn. Mater. {\bf 200}, 110 (1999).

\bibitem{dietlprb97}
T. Dietl, A. Haury, and Y. Merle d'Aubign\'e,
Phys. Rev. B {\bf 55}, R3347 (1997).

\bibitem{jungwirthprb99}
T. Jungwirth, W.A. Atkinson, B.H. Lee, and A.H. MacDonald,
Phys. Rev. B {\bf 59}, 9818 (1999).

\bibitem{bookchapter}
J. K\"{o}nig, J. Schliemann, T. Jungwirth,
and A.H. MacDonald, e-print [http://arXiv.org/abs/cond-mat/0111314],
to be published by Springer Verlag.

\bibitem{schliemannprb01}
J. Schliemann, J. K\"{o}nig, and A.H. MacDonald,
Phys. Rev. B {\bf 64}, 165201 (2001).

\bibitem{msweb}
We have placed a large database with theoretical predictions 
based on the type of model
studied here at {\tt http://unix12.fzu.cz/ms/index.php}.

\bibitem{konigprl00}
J. K\"onig, H.H. Lin, and A.H. MacDonald,
Phys. Rev. Lett. {\bf 84}, 5628 (2000).

\bibitem{schliemannapl01}
J. Schliemann, J. K\"{o}nig, H.H. Lin, and A.H. MacDonald,
Apl. Phys. Lett. {\bf 78}, 1550 (2001).

\bibitem{konigprb01}
J. K\"onig, T. Jungwirth, and A.H. MacDonald,
Phys. Rev. B {\bf 64}, 184423 (2001).

\bibitem{abolfathprb01}
M. Abolfath, T. Jungwirth, J. Brum, and A.H. MacDonald,
Phys. Rev. B {\bf 63}, 054418 (2001).

\bibitem{iiiv}
I. Vurgaftman, J.R. Meyer, and L.R. Ram-Mohan,
J. Appl. Phys. {\bf 89}, 5815 (2001).

\bibitem{gergely0108477}
Z. Gergely and B. Janko, 
e-print [http://arXiv.org/abs/cond-mat/0108477],

\bibitem{dietlrapid01}
T. Dietl, J. K\"{o}nig, and A.H. MacDonald, Phys. Rev. B {\bf 64}, 241201(R) 
(2001).
\end{references}
\end{document}